\documentclass[%
reprint,
superscriptaddress,
frontmatterverbose, 
amsmath,amssymb,aps,
]{revtex4-2}

\usepackage{xcolor}
\usepackage{graphicx}
\usepackage{dcolumn}
\usepackage{bm}
\usepackage{hyperref}
\usepackage[mathlines]{lineno}
\usepackage{amsmath}
\usepackage{multirow}
\usepackage{graphicx}
\usepackage{tabularx}
\usepackage{setspace}
\usepackage{makecell}
\usepackage{natbib}
\usepackage{ctex}
\ctexset{
  figurename = Figure,
  tablename  = Table
}

\usepackage[
scale=0.85, marginratio={1:1, 2:3}, ignoreall,
margin=0.65in,
]{geometry}

\begin{document}

\preprint{APS/123-QED}

\title{$\mu T^2$-NMR: Micro-Scale Correlation Relaxometry for \textit{in-situ} High-Pressure Nuclear Magnetic Resonance}

\newcommand{\equalcontribtext}{These authors contributed equally to this work.}

\author{Thomas Meier}
\email{thomas.meier@sharps.ac.cn}
\affiliation{Shanghai Key Laboratory MFree, Institute for Shanghai Advanced Research in Physical Sciences, Pudong, Shanghai, 201203}
\affiliation{Center for High-Pressure Science and Technology Advance Research, Beijing, China}

\author{Meng Yang (杨梦)}
\altaffiliation{\equalcontribtext}
\affiliation{Center for High-Pressure Science and Technology Advance Research, Beijing, China}

\author{Yishan Zhou (周奕杉)}
\altaffiliation{\equalcontribtext}
\affiliation{Center for High-Pressure Science and Technology Advance Research, Beijing, China}

\author{Yunhua Fu (付云华)}
\affiliation{Center for High-Pressure Science and Technology Advance Research, Beijing, China}
\affiliation{School of Earth and Space Sciences, Peking University, Beijing, China}

\author{Rui Zhang (张锐)}
\affiliation{Center for High-Pressure Science and Technology Advance Research, Beijing, China}
\affiliation{School of Earth and Space Sciences, Peking University, Beijing, China}

\author{Ziliang Wang (王梓良)}
\affiliation{Center for High-Pressure Science and Technology Advance Research, Beijing, China}

\author{Tianyao Zheng (郑天尧)}
\affiliation{Center for High-Pressure Science and Technology Advance Research, Beijing, China}

\author{Rajesh Jana}
\affiliation{Center for High-Pressure Science and Technology Advance Research, Beijing, China}

\author{Takeshi Nakagawa}
\affiliation{Center for High-Pressure Science and Technology Advance Research, Beijing, China}
\date{\today}

\begin{abstract}
Over the last decade, frequency-domain \textit{in-situ} high-pressure nuclear magnetic resonance (NMR) spectroscopy in diamond anvil cells (DACs) has been employed as a structural and electronic probe of condensed matter systems at pressures well into the megabar range. However, extensive spin interactions and sample heterogeneities under pressure often lead to significant spectral overlap, inhibiting independent observation of chemically similar spin sub-species in the same sample. In this work, we introduce a time-domain relaxometry framework specifically suited for DAC experiments, named $\mu T^2$-NMR. Experimental flexibility and operational robustness are benchmarked on three hydrogen-rich molecular solids at pressures up to 72 GPa. We demonstrate that $\mu T^2$-NMR can resolve individual molecular subunits in relaxation space, paving the way for novel high-pressure, high-resolution NMR applications in molecular solids.         
\end{abstract}

\maketitle

\section{Introduction}
High-pressure nuclear magnetic resonance (NMR) spectroscopy has developed into a useful local probe for materials under extreme compression, enabling direct observation of volatile light elements and local electronic environments in diamond anvil cells (DACs)\cite{Eremets1996, Yarger1995, Okuchi2012}. Recent advances in resonator designs \cite{Meier2017} and probe engineering \cite{Meier2019b} have made it possible to perform \textit{in-situ} NMR experiments on sub-nanoliter samples at pressures exceeding 100~GPa and temperatures above 1000~K, opening new experimental opportunities for condensed-matter physics, chemistry, and planetary science \cite{Meier2018ab, Meier2019a, Meier2022b}.
\begin{figure*}[htb]
\includegraphics[width=1\textwidth]{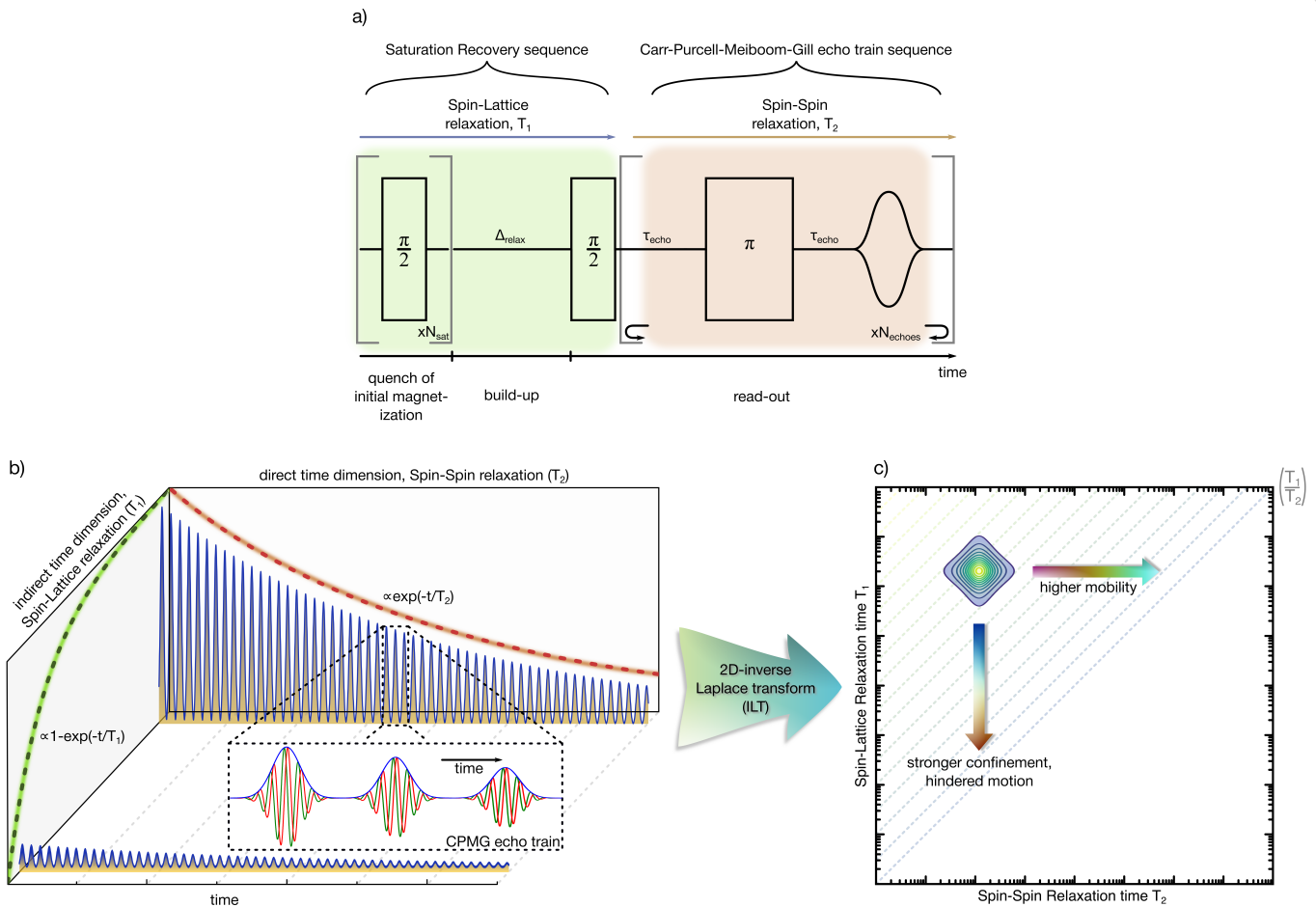}
\caption{\textbf{Schematic representation of $\mu T^2$-NMR pulse scheme and experiment.} \textbf{a)} After an initial quench of z-magnetisation using a string of $\pi/2$-pulses of alternating phases, the recovered magentisation and thus $T_1$, encoded in the build up time $\Delta_{relax}$, is read out by an excitation $\pi/2$-pulse transferring magnetisation into the x-y plane. Subsequent inversion pulses stimulate spin echo build up (CPMG echo trains), encoding $T_2$ relaxation. \textbf{b)} The resulting two-dimensional data set consists of N CPMG echo trains in the time domain. Green and red lines are guide to the eye to illustrate the general time evolution in both dimensions. \textbf{c)} Subsequent Inverse Laplace Transform of the time domain data set results in $T^2$ correlation maps with distribution signals located in the upper diagonal space of the $T_1-T_2$ phase space. The main diagonal ($T_1=T_2$) represents BPP minimum at $\omega_0 \tau_c=1$. Colored arrows indicate the general trend for stronger confinement or hindered motions and higher mobility. For more detail, see text.}
\label{fig:general}
\end{figure*}
\\
To date, most high-pressure NMR studies in DACs have focused on observables probing the magnitude of local magnetic fields $\mathbf{B}_{\mathrm{loc}}(t)$ in the frequency domain, most notably resonance shifts and linewidths. These quantities provide valuable information on electronic structure, bonding environments, and static local fields \cite{Slichter1978,Levitt2000}, and, by extension, on the quantification of spin concentrations \cite{Meier2022b}. They have enabled key insights into hydrogen-rich systems, metal hydrides, and dense molecular solids under extreme compression \cite{Zhou2025Diffusion-drivenConditions, Meier2020}. 
\\
Under high-pressure conditions, however, frequency-domain spectra are often strongly broadened by dipolar interactions, chemical shift dispersion, and sample heterogeneity, resulting in overlapping resonances that complicate spectral interpretation which could be partially mitigated using static line-narrowing techniques, leading to substantially improved spectral resolution in selected hydrogen-based systems \cite{Meier2019,Meier2021a,Yang2024HexagonalNMR}. While such advances extend the applicability of frequency-domain NMR under extreme conditions, they primarily enhance access to static or quasi-static spectral information.
\\
By contrast, frequency-domain observables remain intrinsically less sensitive to dynamical heterogeneity and molecular motion when distinct environments produce similar local field distributions. In such cases, markedly different relaxation pathways may coexist despite nearly indistinguishable resonance frequencies. As a result, frequency-domain NMR alone provides only limited access to molecular dynamics and microscopic heterogeneity in many compressed materials, motivating the use of relaxation-based approaches that directly probe time-dependent fluctuations of $\mathbf{B}_{\mathrm{loc}}(t)$.
\\
Nuclear spin relaxation is intrinsically sensitive to dynamical processes and local fluctuations of magnetic fields. Spin-lattice ($T_1$) and spin-spin ($T_2$) relaxation times vary over several orders of magnitude depending on the nature and timescale of molecular motion, electronic interactions, and lattice dynamics \cite{Redfield1957OnProcesses}. 
In extreme cases, relaxation times range from microseconds in strongly interacting or metallic systems \cite{Meier2016} to hours or longer in rigid or dilute spin environments \cite{Reynhardt19973CDiamond}. Importantly, relaxation pathways may differ substantially even when static NMR spectra appear similar or unresolved, making relaxation a particularly powerful probe in systems where frequency-domain information is ambiguous.
\\
Accessing relaxation-derived information in heterogeneous systems generally requires moving beyond single-parameter descriptions or one-dimensional experimental set-ups. Two-dimensional relaxation correlation experiments provide a natural framework in this regard. Correlating spin--lattice ($T_1$) and spin--spin ($T_2$) relaxation times, relaxation maps—hereafter referred to as $T^2$ maps—allow distinct dynamical regimes and coexisting environments to be separated in relaxation space, even when their spectral signatures overlap. Such approaches are well established in low-field and benchtop NMR, where they have been employed to characterize heterogeneous systems, diffusion processes, and molecular mobility across a wide range of length and time scales \cite{Song2002,Galvosas2007OnMaterials}. Their application to \textit{in-situ} high-pressure NMR experiments in diamond anvil cells, however, has so far remained largely unexplored.
\\
In this work, we introduce $\mu T^2$-NMR, a micro-scale $T_1$--$T_2$ correlation relaxometry framework specifically adapted to \textit{in-situ} high-pressure NMR experiments in DACs. By combining saturation-recovery encoding\cite{Bloembergen1948} of $T_1$ with Carr--Purcell--Meiboom--Gill (CPMG)\cite{Carr1954EffectsExperiments, Meiboom1958ModifiedTimes} detection of $T_2$, and employing two-dimensional inverse Laplace transformation, correlated relaxation-time distributions can be recovered from sub-nanoliter samples under extreme compression, see figure \ref{fig:general}. We benchmark the method on hydrogen-rich molecular solids and evaluate its operational stability, signal-to-noise behavior, and thermal robustness under high-pressure conditions. The results demonstrate that $\mu T^2$-NMR enables the resolution of distinct motional regimes and coexisting local environments beyond the limits of conventional frequency-domain NMR, establishing relaxation correlation spectroscopy as a practical and broadly applicable tool for probing dynamics in materials under extreme conditions.

\section{Theoretical background}
\subsection{Spin-Lattice and Spin-Spin Relaxation} 
Nuclear spin relaxation arises from time-dependent fluctuations of the local magnetic field experienced by the nuclear spins,
$\mathbf{B}_{\mathrm{loc}}(t)$, which may originate from molecular motion, lattice vibrations, electronic interactions, and mutual spin--spin couplings. In a strong external magnetic field $B_0 \mathbf{e_z}$, it is convenient to decompose these fluctuations into components perpendicular and parallel to the quantization axis,
\begin{equation}
\mathbf{B}_{\mathrm{loc}}(t) =
B_{\perp}(t)(\mathbf{e}_x+\mathbf{e}_y) + B_{\parallel}(t)\mathbf{e}_z.
\end{equation}
The perpendicular component ${B}_{\perp}(t)$ induces transitions between Zeeman levels and thus governs energy exchange with the surrounding lattice, while the parallel component $B_{\parallel}(t)$ modulates the Larmor precession frequency and contributes to dephasing.
\\
Within Redfield theory, relaxation rates are determined by the spectral densities $J(\omega)$ of these fluctuating field components \cite{Bloembergen1948,Redfield1957OnProcesses,Slichter1978}, and are defined as the Fourier transform of the corresponding autocorrelation functions of the stationary fluctuating local fields,
\begin{equation}
J_{\alpha}(\omega) =
\int_{\mathbb{R}}
\langle B_{\alpha}(t) B_{\alpha}(0) \rangle
e^{-i\omega t} \, \mathrm{d}t,
\qquad
\alpha \in \{\perp,\parallel\}.
\end{equation}

For a spin-$1/2$ nucleus, the longitudinal and transverse relaxation rates may then be written, to leading order, as
\begin{align}
\frac{1}{T_1} &\propto J_{\perp}(\omega_0), \\
\frac{1}{T_2} &\propto \frac{1}{2}J_{\perp}(\omega_0) + J_{\parallel}(0),
\end{align}
where $\omega_0$ is the Larmor frequency. While the proportionality constants depend on the specific interaction mechanism, these expressions emphasize that $T_1$ and $T_2$ probe distinct frequency components and polarization channels of the same underlying fluctuating fields.

In liquids undergoing isotropic, fast molecular motion, the distinction between $J_{\perp}$ and $J_{\parallel}$ is often of limited practical relevance, as motional averaging leads to similar spectral densities across components. In solids and partially ordered systems, however, this separation becomes essential. Under high pressure, molecular reorientation is frequently hindered, anisotropic interactions persist, and slow or quasi-static fluctuations contribute significantly to $J_{\parallel}(0)$.

As a result, $T_2$ relaxation in compressed solids is often dominated by low-frequency or static contributions arising from dipolar couplings, structural disorder, or slowly evolving local environments, whereas $T_1$ remains governed by higher-frequency fluctuations capable of inducing Zeeman transitions. Pressure-induced changes in bonding topology, hydrogen-bond symmetrization, or partial proton mobility may therefore affect $T_1$ and $T_2$ in qualitatively different ways, even when originating from the same microscopic degrees of freedom.

This separation is particularly relevant for hydrogen-rich systems in diamond anvil cells, where strong dipolar couplings, anisotropic confinement, and restricted motion persist over wide pressure ranges. In such cases, the relative magnitudes and pressure evolution of $J_{\perp}(\omega_0)$ and $J_{\parallel}(0)$ encode complementary information about molecular dynamics, confinement, and local structural heterogeneity.

For a single exponential correlation function characterized by a correlation time $\tau_c$, the spectral density assumes a Lorentzian form,
\begin{equation}
J_{\alpha}(\omega) =
\frac{2 \langle B_{\alpha}^2 \rangle \tau_c}
{1 + \omega^2 \tau_c^2}.
\end{equation}
In this simplified picture, $T_1$ relaxation is most efficient when $\omega_0 \tau_c \sim 1$, whereas $T_2$ relaxation receives additional contributions from slow motions with $\omega \tau_c \ll 1$. 
In real materials, particularly under extreme compression, relaxation is rarely governed by a single correlation time or a unique relaxation constant. Instead, spatial and dynamical heterogeneity give rise to non-exponential magnetization recovery and decay, reflecting the coexistence of multiple relaxation pathways associated with distinct local environments or motional regimes. One-dimensional relaxation experiments can capture such behavior phenomenologically through stretched or multi-component decays, but they do not, in general, preserve information on how longitudinal and transverse relaxation processes are correlated within the same microscopic environments.
\\
In heterogeneous or dynamically complex systems, distinct environments may exhibit similar resonance frequencies while possessing markedly different combinations of $J_{\perp}$ and $J_{\parallel}$. Conversely, multiple relaxation components may coexist within a single spectral line, reflecting variations in molecular mobility, confinement, or bonding that remain invisible in the frequency domain. For example, amide ($\mathrm{NH}_2$) and imide ($\mathrm{NH}$) protons frequently overlap in one-dimensional $^1$H NMR spectra in the absence of line-narrowing techniques, as their chemical shifts—i.e.\ the static contribution to $\mathbf{B}_{\mathrm{loc}}$—are often comparable. In contrast, their spin--lattice relaxation times may differ by up to two orders of magnitude, reflecting distinct local dynamics and coupling mechanisms. A joint description of longitudinal and transverse relaxation, therefore, provides a more complete representation of the underlying dynamics than either parameter alone.

\subsection{Inverse Laplace Transform, kernel formulation, and statistical robustness}

Relaxation correlation experiments yield a two-dimensional time-domain data set in which the experimentally measured signal depends on both the longitudinal recovery delay and the transverse echo time. Recovering correlated $T_1$--$T_2$ information from such data requires solving an inverse Laplace problem, in which the measured signal is represented as a superposition of exponential relaxation modes, see figure \ref{fig:general}b. In its continuous form, this inversion constitutes a Fredholm integral equation of the first kind, which is intrinsically ill-posed: small perturbations of the input data, such as experimental noise, can lead to large variations in the recovered relaxation distributions, and unique solutions do not generally exist \cite{Song2002,Galvosas2007OnMaterials}.

Formally, the measured signal $S(t_{\mathrm{SR}},t_{\mathrm{CPMG}})$ may be written as:
\begin{equation}
\begin{aligned}
S(t_{\mathrm{SR}},t_{\mathrm{CPMG}})
&=
\iint_{\mathbb{R}}
P(T_1,T_2)\,
K_{T_1}(t_{\mathrm{SR}},T_1) \\
&\qquad\qquad\times
K_{T_2}(t_{\mathrm{CPMG}},T_2)\,
\mathrm{d}T_1\,\mathrm{d}T_2 ,
\end{aligned}
\end{equation}
where $P(T_1,T_2)$ denotes the joint relaxation-time distribution and
$K_{T_1}$ and $K_{T_2}$ are the longitudinal recovery and transverse decay kernels, respectively.
\\
\begin{align}
K_{T_1}(t_{\mathrm{SR}},T_1) &= 1 - \exp\!\left(-\frac{t_{\mathrm{SR}}}{T_1}\right), \\
K_{T_2}(t_{\mathrm{CPMG}},T_2) &= \exp\!\left(-\frac{t_{\mathrm{CPMG}}}{T_2}\right),
\end{align}
although deviations from these idealized expressions may arise from experimental imperfections or non-ideal pulse behavior.
\\
In practice, the continuous Fredholm formulation is approximated by discretizing the $(T_1,T_2)$ relaxation space onto a finite grid of logarithmically spaced values. The integral equation may then be expressed in matrix form as
\begin{equation}
\mathbf{S} = \mathbf{K}_{T_1}\, \mathbf{P}\, \mathbf{K}_{T_2}^{\mathrm{T}},
\end{equation}
where $\mathbf{S}$ is the measured signal matrix, $\mathbf{K}_{T_1}$ and $\mathbf{K}_{T_2}$ are the discretized kernel matrices, and $\mathbf{P}$ represents the two-dimensional relaxation-time distribution. Each element of $\mathbf{P}$ corresponds to the contribution of spins characterized by a specific $(T_1,T_2)$ pair.
\\
To obtain reproducible and physically meaningful solutions, the inversion is carried out under explicit regularization and non-negativity constraints. In this work, a Tikhonov-regularized non-negative least-squares approach is employed, minimizing the functional
\begin{equation}
\left\| \mathbf{S}_{\mathrm{exp}} - \mathbf{K}_{T_1} \mathbf{P} \mathbf{K}_{T_2}^{\mathrm{T}} \right\|^2
+ \lambda \left\| \mathbf{L} \mathbf{P} \right\|^2,
\end{equation}
where $\mathbf{S}_{\mathrm{exp}}$ denotes the experimental data, $\lambda$ is the regularization parameter, and $\mathbf{L}$ is a smoothing operator acting in relaxation space. The regularization term suppresses unphysical oscillations and noise amplification inherent to the inverse Laplace problem, while the non-negativity constraint ensures that the recovered distributions remain physically interpretable.
\\
The inverse Laplace transformation and subsequent analysis are implemented in a custom Python framework developed for this work. The code performs logarithmic discretization of the relaxation axes, kernel construction, regularized inversion, and post-processing of the resulting $T^2$ maps. While the absolute amplitudes of the recovered distributions depend on the choice of discretization and regularization strength, the locations, shapes, and relative separation of distinct relaxation components are robust against reasonable variations of these numerical parameters. Accordingly, the $T^2$ maps presented here are interpreted in terms of correlated relaxation regimes and dynamical heterogeneity rather than as quantitative population measures.
\\
\begin{figure*}[htb]
\includegraphics[width=0.85\textwidth]{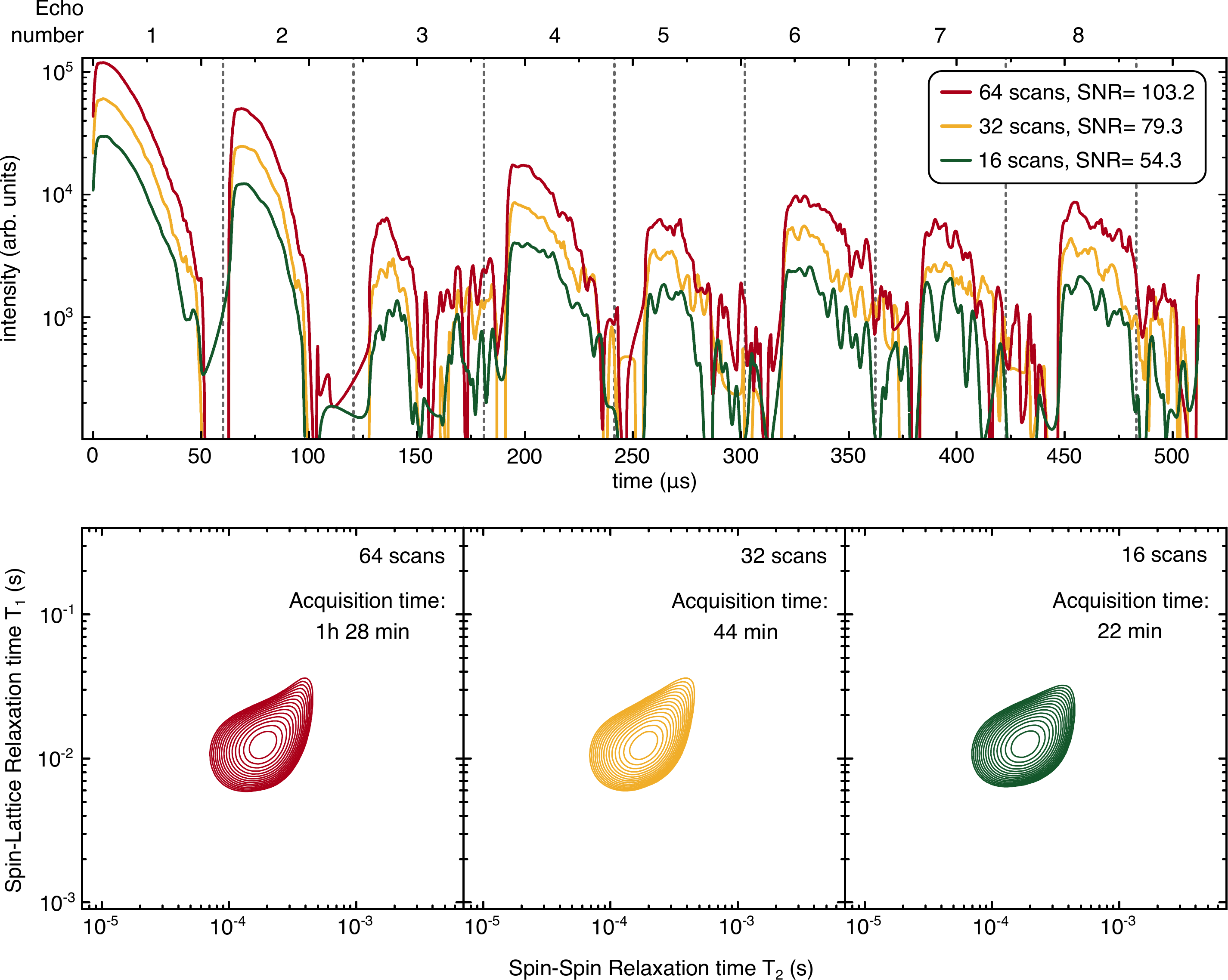}
\caption{\textbf{Comparison of $\mu T^2$-map distributions and time domain signal-to-noise in Melamine at 65 GPa} Top: time domain data of three CPMG echo trains recorded under 64 (red), 32 (yellow), and 16 (green) scan accumulations. only the first 8 echoes are shown. Horizontal markers indicate the accumulation time windows in the CPMG train and the corresponding echoes. Bottom: Respective $\mu T^2$-maps of the amide-$NH_2$ groups in melamine. Total acquisition time is indicated for each distribution map. As can be seen, the data quality of the computed $\mu T^2$-maps remains robust under significant reduction of time domain SNR.}
\label{fig:SNR}
\end{figure*}
To assess the statistical robustness of the recovered relaxation features, a bootstrapping procedure is employed. Synthetic data sets are generated by resampling the experimental time-domain data with replacement and repeating the inversion for each realization. The resulting ensemble of $T^2$ maps provides a statistical measure of the stability of individual relaxation components with respect to experimental noise and finite signal-to-noise ratio. Features that persist across bootstrap realizations are identified as robust, whereas components that vary strongly between realizations are treated with caution. This procedure allows experimentally meaningful relaxation regimes to be distinguished from inversion artifacts without relying on a single nominal solution.
\\
\section{Experimental}
Throughout this work, CuBe-based wide panoramic DACs were used, equipped with 250 $\mu m$ culeted diamonds. Rhenium gaskets were prepared by pre-indenting to about 10-15 $\mu m$ and laser drilling of about 80 $\mu m$ diameter sample cavities, resulting in effective sample volumes of approximately 2.5 pl. Diamond anvils were covered using PVD with a 1 $\mu m$ thick layer of copper, and focused ion beam shaping was subsequently used to shape double-stage Lenz-Lens resonator structures \cite{Meier2018c}. Two RF coils placed around the coated diamond anvils were connected to form a Helmholtz coil arrangement that drives the Lenz-lens micro-resonators.
\\
Samples of Formamide ($HCONH_2$, liquid, 4N), Melamine ($C_3H_6N_6$, powder, 4N) and Trifluoroacetamide (TFA, $CF_3CONH_2$, powder, 4N) were loaded in the DAC's sample chambers. Additionally, samples of TFA and melamine were mixed with fine gold particles (grain size $<1\mu m$) as absorbers for \textit{in-situ} laser heating. Internal pressures were monitored using the first derivative of the line shift of the diamond Raman spectrum\cite{Akahama2004, Akahama2006}.
\\
Cells loaded with melamine and TFA were precompressed to 30-35 GPa and laser heated to about 1200-1500 K in a double-sided laser heating system to access recently discovered nitride compounds such as $\beta-C(NH)_2$ exhibiting intercalating carbon and imide networks\cite{Koller2024SimpleCarbon}. 
\\
All NMR experiments were conducted on a Q-band EPR-electromagnet retrofitted for low-field $^1H$-NMR at 1.2 T field strength. DACs were mounted on home-built NMR probes and driven by Tecmag Redstone spectrometers at hydrogen resonance frequencies of approximately $f_0=51~MHz$. Using pulse nutation experiments, optimal $\pi/2$ pulse lengths of approximately 7-10 $\mu s$ at an average radio-frequency power of 10 W were determined. 
\\
2D-Correlation-Relaxometry experiments, $\mu T^2$-NMR, were conducted using a pulse sequence schematically depicted in figure \ref{fig:general}a. The Saturation recovery pulse train consists of 10 $\pi/2$ pulses with alternating phases, each with a successively smaller separation, to quench the z-magnetization $M_z$ of the hydrogen spin systems. The relaxation separation $\Delta_{relax}$ is varied logarithmically from 0.5 $\mu s$ to 10-42 s, typically with 128 increments. The read-out of the recovered values of $M_z$ is conducted with a Carr-Purcell-Meiboom-Gill (CPMG) echo train of $N_{echoes}$ $\pi$-pulses. The number of CPMG echoes ($N_{echoes}$) is usually spin-system-dependent and is typically chosen to be 64, 128, or 256. The resulting raw 2D time-domain datasets are analyzed using a Python framework, which performs the inverse Laplace transform and post-processing.  
\\
\section{Data quality robustness and acquisition efficiency}
The practical performance of the SRCPMG-based $\mu T^2$ experiment is governed by the signal-to-noise ratio (SNR) of the underlying one-dimensional time-domain data. To assess the robustness of the method under realistic experimental constraints, we compare SRCPMG echo trains recorded with varying numbers of scan accumulations on melamine at 65 GPa, as a representative test system (Fig.~\ref{fig:SNR}).
\\
Figure~\ref{fig:SNR} top shows the first eight echoes of CPMG trains acquired with 64, 32, and 16 scan accumulations, corresponding to total acquisition times of 1~h~28~min, 44~min, and 22~min, respectively. As expected, time-domain SNR decreases systematically with reduced signal averaging. Nevertheless, the overall echo train structure and decay behavior remain clearly resolved even with the fewest accumulations, indicating that the SRCPMG sequence provides sufficient intrinsic sensitivity for successful ILT post-processing.
\\
The corresponding $\mu T^2$ maps derived from these data sets are shown on the bottom of fig.~\ref{fig:SNR}. As shown, despite substantial reductions in acquisition time and time-domain SNR, the recovered relaxation distributions remain remarkably well resolved. The location, overall shape, and orientation of the dominant amide proton relaxation component in melamine are preserved across all three measurements. While a minor broadening of the distribution is observed at lower SNR, no spurious components or qualitative distortions are observed.
\\
These observations demonstrate that the $\mu T^2$-NMR approach is not limited by marginal time-domain sensitivity, but instead remains operationally robust under significant reductions in scan number and total experiment time. In practice, this robustness enables flexible adaptation of the acquisition parameters to experimental constraints imposed by pressure stability, temperature control, or sample lifetime, without compromising the qualitative integrity of the resulting relaxation correlation maps.
\\
\section{Radio-frequency heating and temperature stability}
\begin{figure}[htb]
\includegraphics[width=1\columnwidth]{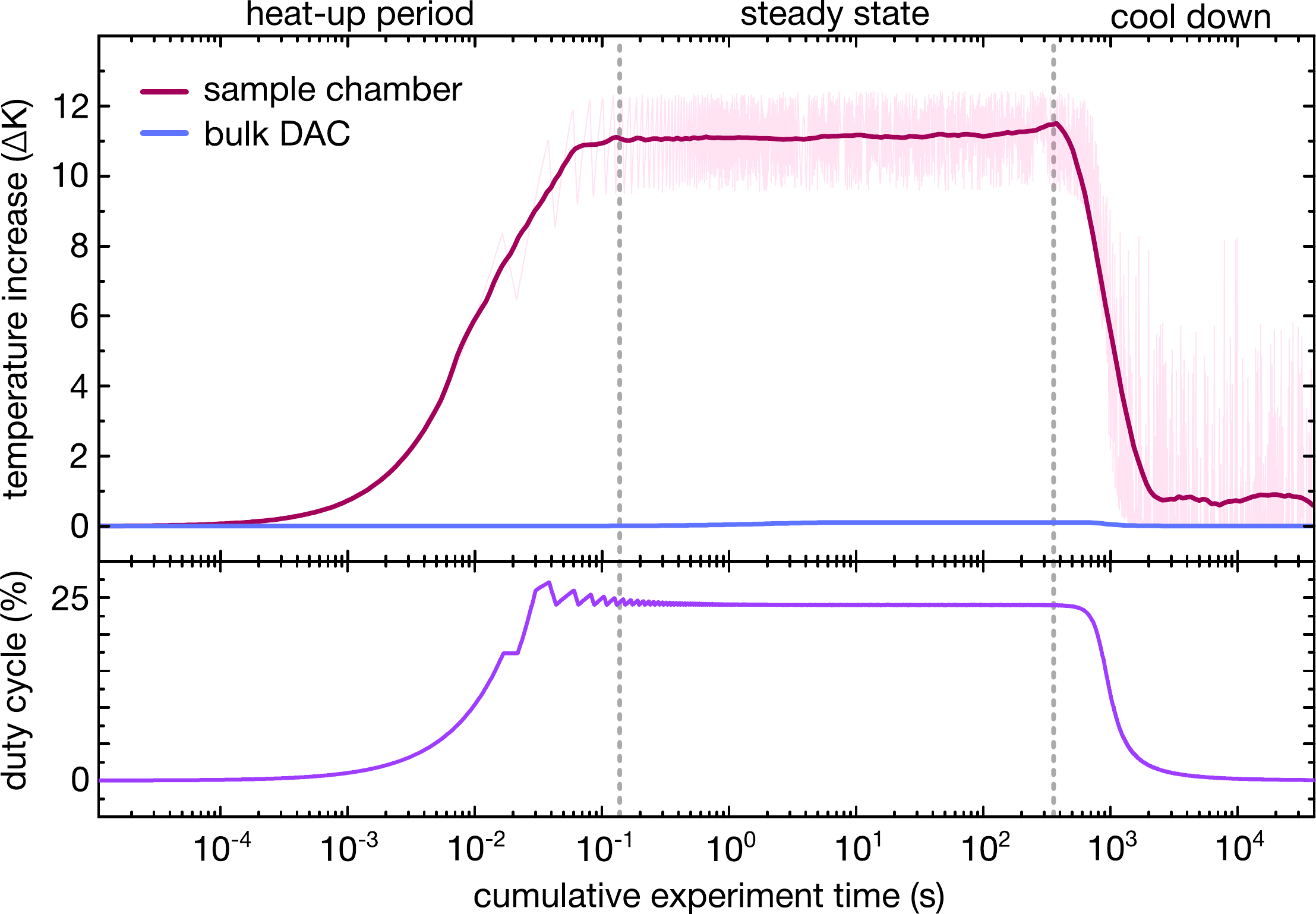}
\caption{\textbf{Estimation of the temperature increase in the sample chamber during SRCPMG pulse experiments.} During the heat-up period, pulse separations are shortest due to short values of $\Delta_{relax}$, leading to significant temperature build-up in the cavity where the RF-magnetic field, generated by the micro-resonators, is most focused. For longer values of $\Delta_{relax}$, the system enters a steady-state where build-up temperature gets efficiently dissipated by the surrounding large heat sinks (diamond anvils, rhenium gasket, etc.). For longer separation times, heat dissipation through the diamonds becomes dominant, and the temperature returns to its initial value. As can be seen, the sample chamber temperature (pink) follows qualitatively the RF duty cycle imposed by the pulse sequence (purple). The blue line depicts the estimated temperature increase in the bulk DAC.
}
\label{fig:temp}
\end{figure}
SRCPMG-based relaxation correlation experiments involve a large number of radio-frequency (RF) pulses applied over extended time periods. For a typical $\mu T^2$ experiment, the cumulative number of $\pi/2$ and $\pi$ pulses easily exceeds $10^5$--$10^6$, with pulse separations that may be as short as a few microseconds during the early stages of the saturation-recovery encoding. Under such conditions, RF-induced heating cannot be neglected. In fact, preliminary tests on larger samples outside diamond anvil cells yielded substantial temperature increases and, in extreme cases, sample melting. This practical limitation likely contributes to the limited adoption of SRCPMG-type experiments in conventional solid-state NMR, particularly for hydrogen-rich systems.
\\
To assess the thermal load under realistic high-pressure conditions, the temperature evolution during SRCPMG pulsing was estimated using a simplified two-node thermal model. In this approach, the sample chamber and the surrounding DAC assembly are treated as coupled thermal reservoirs. RF power dissipation is assumed to occur predominantly within the micro-resonator and sample volume, where the RF magnetic field is most strongly concentrated. In contrast, heat dissipation proceeds through the surrounding diamond anvils, metallic gasket, and backing plates, which act as large thermal sinks. The instantaneous heating rate is governed by the RF duty cycle imposed by the pulse sequence, whereas cooling is described by effective thermal coupling between the two nodes.
\\
Figure~\ref{fig:temp} summarizes the resulting temperature estimates for a representative SRCPMG experiment. For this simulation, 256 scan accumulations and corresponding relaxation-separation increments $\Delta_{\mathrm{relax}}$ were assumed, with a maximum recovery delay of 10~s. The number of echoes per CPMG train was fixed to 128, using 10~$\mu$s $\pi/2$ pulses at an average RF pulse power of 50~W. These parameters were deliberately chosen to represent a borderline case to emphasize RF-induced heating effects. In practice, pulse sequence parameters are selected more conservatively to mitigate heating, resulting in lower absolute temperature increases than those shown in Fig.~\ref{fig:temp}, while preserving the same qualitative behavior.
\\
During the initial heat-up period, short pulse separations associated with small values of the recovery delay $\Delta_{\mathrm{relax}}$ lead to a rapid increase in RF duty cycle and a corresponding rise in the sample chamber temperature. As the experiment progresses to longer recovery delays, the duty cycle decreases, and a dynamic steady state is reached, in which RF heating is approximately balanced by heat dissipation into the surrounding DAC components. With further increases in pulse separations, cooling dominates, and the sample temperature relaxes back toward its initial value.
\\
Several key observations emerge from this analysis. First, although transient temperature increases of several kelvin can occur in the sample chamber, the temperature of the bulk DAC remains essentially unchanged, reflecting the large thermal mass and efficient heat sinking provided by the diamond anvils. Second, the temporal evolution of the sample temperature closely follows the RF duty cycle imposed by the pulse sequence, confirming that heating is dominated by cumulative RF power rather than by individual pulses. Finally, no runaway heating is observed under typical $\mu T^2$ acquisition conditions, indicating that SRCPMG experiments can be performed in DACs with acceptable thermal stability.
\\
These results highlight a significant and somewhat counterintuitive advantage of high-pressure DAC environments. While RF heating poses a severe limitation for SRCPMG experiments on macroscopic samples, the extreme confinement and excellent thermal coupling in DACs effectively suppress excessive temperature excursions. As a result, $\mu T^2$-NMR experiments can be carried out safely and reproducibly under high-pressure conditions, even for hydrogen-rich systems subjected to intense RF pulse trains.

\section{Spectral Overlap and Contrast in Relaxation Space}
\begin{figure*}[htb]
\includegraphics[width=.9\textwidth]{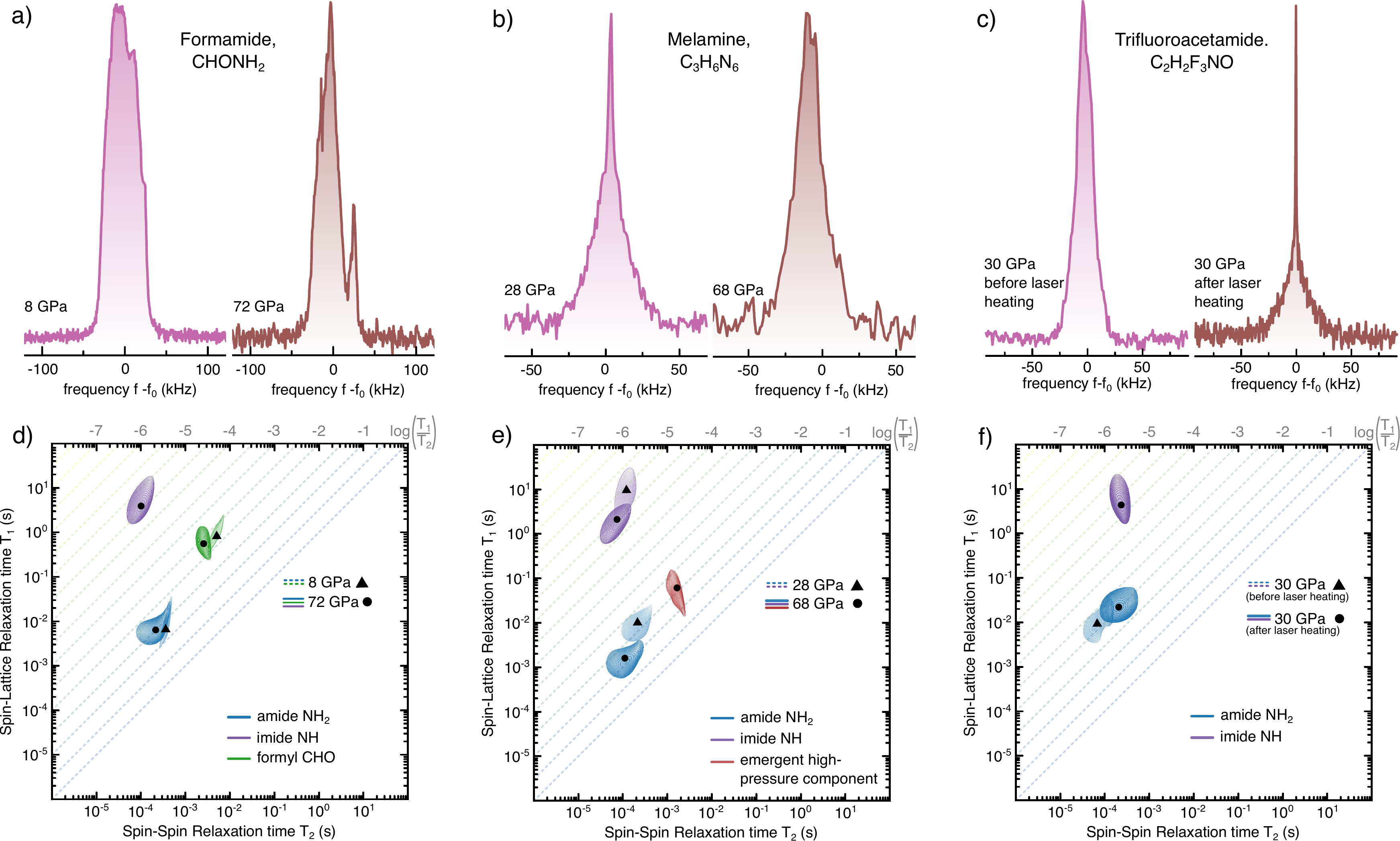}
\caption{\textbf{Comparison of frequency-domain $^1$H-NMR spectra and corresponding $\mu T^2$ maps for molecular solids in diamond anvil cells.}
(\textbf{a--c}) Representative one-dimensional $^1$H-NMR spectra of formamide (CHONH$_2$), melamine (C$_3$H$_6$N$_6$), and trifluoroacetamide (C$_2$H$_2$F$_3$NO) at selected pressures. In all three systems, strong spectral overlap is observed in the frequency domain, arising primarily from amide and imide proton environments and, in the case of formamide, additional formyl (CHO) protons. With increasing pressure or thermal treatment, new overlapping spectral features emerge that cannot be reliably separated based on frequency-domain information alone. (\textbf{d--f}) Corresponding $\mu T^2$ correlation maps acquired under identical conditions. Despite the severe spectral overlap in the one-dimensional spectra, distinct relaxation components are clearly resolved in relaxation space. (\textbf{d}) For formamide, overlapping amide (NH$_2$) and formyl (CHO) proton resonances at low pressure give rise to separate relaxation distributions. At higher pressure, an additional relaxation component appears, likely associated with the formation of imide (NH) protons, indicating pressure-induced chemical transformation. (\textbf{e}) In melamine, laser heating at 30~GPa leads to the formation of $\beta$-C(NH)$_2$, accompanied by a new relaxation component attributed to imide NH groups. At pressures above 45~GPa, a third relaxation pathway becomes apparent, clearly distinguishable in the $\mu T^2$ maps despite remaining spectrally unresolved. (\textbf{f}) For trifluoroacetamide, the frequency-domain spectra are dominated by overlapping amide NH$_2$ resonances, whereas the corresponding $\mu T^2$ maps reveal multiple relaxation components. After laser heating at 30~GPa, additional relaxation pathways emerge with characteristics comparable to imide NH groups. Structural identification of these high-pressure phases is currently ongoing.
}
\label{fig:spectra}
\end{figure*}
One of the central limitations of high-pressure $^1$H-NMR in diamond anvil cells is the severe spectral congestion encountered in the frequency domain. Strong homonuclear dipolar couplings, pressure-induced disorder, and chemical shift dispersion frequently lead to broad and strongly overlapping resonances, particularly in hydrogen-rich molecular solids. As a result, chemically and structurally distinct proton environments may become indistinguishable in one-dimensional spectra, even when their underlying dynamics differ substantially. This limitation becomes increasingly pronounced under higher pressures and temperatures, precisely in the regimes where structural and dynamic information is of great interest.
\\
$\mu T^2$ correlation experiments presented here address this problem by utilizing contrast in relaxation behavior rather than relying on spectral separation. Figure~\ref{fig:spectra} compares conventional frequency-domain $^1$H-NMR spectra with the corresponding $\mu T^2$-maps for three distinct molecular solids -- formamide, melamine, and trifluoroacetamide -- under high-pressure conditions. In all cases, one-dimensional $^1H$-NMR spectra exhibit overlapping contributions from different molecular sub-units, precluding reliable spectral decomposition. By contrast, the corresponding relaxation maps reveal well-separated distributions in $(T_1,T_2)$ space, enabling distinct molecular subunits to be resolved despite the absence of spectral resolution.
\\
Formamide provides an instructive example, fig \ref{fig:spectra}a and d. At low pressure, the $^1$H-NMR spectrum consists of at least two strongly overlapping contributions arising from amide ($NH_2$) and formyl ($CHO$) protons. While these environments cannot be disentangled spectrally in the absence of line-narrowing techniques, the corresponding $\mu T^2$ maps exhibit clearly separated relaxation distributions, reflecting their distinct local dynamics and coupling mechanisms. Upon further compression, an additional relaxation component emerges that is not readily identifiable in the frequency domain. The appearance of this third distribution in relaxation space is consistent with the formation of imide ($NH$) protons and indicates a pressure-induced chemical transformation that would be difficult to detect from spectral data alone.
\\
A similar contrast between spectral overlap and relaxation-space separation is observed for melamine (see fig \ref{fig:spectra}b and e). At moderate pressures, frequency-domain spectra are dominated by broad, overlapping features associated with amide $NH_2$ groups. Laser heating at 30~GPa initiates the formation of $\beta$-C(NH)$_2$ \cite{Koller2024SimpleCarbon}, accompanied by the appearance of a new relaxation component attributed to imide NH protons. At pressures above approximately 45~GPa, a third relaxation pathway becomes apparent in the $\mu T^2$ maps, which is not directly observable in the frequency spectra at those pressures.
\\
Trifluoroacetamide exhibits an analogous behavior, see fig. \ref{fig:spectra}c and f. Its $^1$H-NMR spectra are dominated by overlapping amide resonances across the investigated pressure range, providing little direct insight into changes in molecular dynamics or chemical environment. In contrast, the $\mu T^2$ maps reveal multiple relaxation components occupying distinct regions of relaxation space. Following laser heating at 30~GPa, additional relaxation pathways emerge, with characteristics comparable to those of imide NH groups, indicating the formation of new local proton environments. While the detailed structural assignment of these high-pressure phases remains under investigation, their presence is unambiguously captured in relaxation space.
\\
These examples demonstrate that relaxation correlation spectroscopy provides a powerful and complementary perspective to conventional frequency-domain NMR under extreme conditions. Whereas spectral overlap severely limits the interpretability of one-dimensional $^1$H spectra in dense hydrogenous materials, $\mu T^2$ maps retain sensitivity to dynamical heterogeneity and coexisting environments. Importantly, this contrast is achieved without requiring spectral resolution or line-narrowing techniques, making $\mu T^2$-NMR particularly well suited for \textit{in-situ} studies of molecular solids in diamond anvil cells.

\section{Conclusion}
In this work, we have introduced $\mu T^2$-NMR as a micro-scale relaxation correlation framework tailored to \textit{in-situ} high-pressure NMR experiments in diamond anvil cells. By combining saturation-recovery encoding of $T_1$ with CPMG-based detection of $T_2$, and by reconstructing correlated relaxation-time distributions via two-dimensional inverse Laplace transformation, $\mu T^2$-NMR enables access to molecular dynamics and microscopic heterogeneity that remain largely inaccessible in conventional frequency-domain experiments. Using hydrogen-rich molecular solids as representative test systems, we have demonstrated that severe spectral overlap in one-dimensional $^1$H-NMR spectra can be overcome by exploiting contrast in relaxation space, allowing distinct motional regimes and coexisting local environments to be resolved under extreme compression.
\\
Beyond qualitative visualization, the $\mu T^2$ framework provides access to a set of quantitative descriptors extracted directly from the relaxation maps. Within the accompanying Python-based analysis pipeline, parameters such as the center of gravity in $(T_1,T_2)$ space, the minimum full width at half maximum (MFWHM), principal axis lengths, and orientation of relaxation distributions can be systematically evaluated. These observables enable a compact and reproducible description of pressure- and temperature-induced changes in relaxation behavior, facilitating comparative studies across different materials, loading conditions, or experimental protocols. While absolute signal amplitudes in $T^2$ maps are not interpreted quantitatively, the relative positions, shapes, and separations of relaxation components provide robust and physically meaningful metrics of dynamical evolution.
\\
At the same time, several practical limitations of the method must be acknowledged. The accessible dynamical range in both $T_1$ and $T_2$ is constrained by experimental timing, signal-to-noise ratio, and RF duty cycle. In particular, systems with extremely short transverse relaxation times, for which spin echoes cannot be efficiently refocused, remain inaccessible to SR--CPMG-based correlation experiments. Similarly, relaxation components lying far outside the experimentally sampled time windows cannot be reliably recovered, placing intrinsic bounds on the observable relaxation spectrum. These limitations are not unique to $\mu T^2$-NMR but are inherent to relaxation correlation methods more generally and must be considered when designing experiments and interpreting results.
\\
Thermal stability represents an additional experimental consideration. As demonstrated here, SR--CPMG experiments involve very large numbers of RF pulses applied at short separations, which can lead to significant local heating under unfavorable conditions. While the confined geometry and efficient heat dissipation in diamond anvil cells mitigate these effects compared to bulk samples, careful optimization of pulse parameters remains essential, particularly for temperature-sensitive systems. The ability to model and estimate RF-induced heating, as implemented in this work, therefore constitutes an important practical component of the methodology.
\\
Despite these constraints, $\mu T^2$-NMR is broadly applicable to a wide range of condensed-matter systems accessible in DACs, including molecular solids, hydrogen-rich compounds, and materials undergoing pressure- or temperature-driven transformations. The method is compatible with existing high-pressure NMR hardware and does not rely on spectral resolution or line-narrowing techniques, making it particularly attractive for systems dominated by strong spin interactions or structural disorder. Looking forward, extensions to other relaxation correlation schemes, multi-nuclear experiments, or combined pressure–temperature protocols appear feasible and may further expand the scope of relaxation-based probes under extreme conditions.
\\
Overall, $\mu T^2$-NMR establishes relaxation correlation spectroscopy as a practical and informative complement to frequency-domain NMR in diamond anvil cells. By shifting the focus from spectral separation to dynamical contrast, the approach opens new experimental pathways for investigating molecular motion, bonding rearrangements, and heterogeneous dynamics in materials subjected to extreme compression.
\\
Beyond methodological development, $\mu T^2$-NMR correlation relaxometry offers a new experimental pathway for probing hydrogen-bearing materials under conditions relevant to Earth and planetary interiors\cite{Hu2021}.
Many geophysically critical phases—such as hydrous silicates, dense molecular solids, hydrogen-rich organics, and planetary ice analogues—exhibit complex, heterogeneous hydrogen dynamics that are difficult to probe with conventional spectroscopic methods at high pressure \cite{Ji2020a, Yuan2017}. By separating distinct dynamical regimes in $(T_1,T_2)$ space rather than relying on spectral resolution or absolute signal amplitudes, $\mu T^2$-NMR enables the identification of coexisting proton environments associated with hydrogen bonding, molecular reorientation, diffusion, or partial sublattice mobility within a single compressed sample.
\\
This capability is particularly relevant for studies of hydrogen transport, proton disorder, and superionic or pre-superionic behavior in deep-Earth and planetary materials, where subtle changes in hydrogen mobility can strongly influence electrical conductivity, rheology, and thermal transport \cite{Mao2020}.
\\
More generally, correlation relaxometry provides a framework for connecting microscopic hydrogen dynamics to macroscopic physical properties across pressure–temperature regimes that remain challenging for diffraction- or transport-based techniques.
As high-pressure NMR sensitivity and temperature control continue to improve, $\mu T^2$-NMR has the potential to become a broadly applicable tool for exploring hydrogen-mediated processes in geoscience, planetary science, and related fields.

\begingroup
\renewcommand\thefootnote{*}
\footnotetext{These authors contributed equally to this work.}
\endgroup

\section*{Acknowledgements}

This work was supported by the National Science Foundation of China (W2432007, 42150101), the National Key Research and Development Program of China (2022YFA1402301) as well as the Shanghai Key Laboratory for Novel Extreme Condition Materials, China (no. 22dz2260800), Shanghai Science and Technology Committee, China (no. 22JC1410300).

\section*{Author Contributions Statement}
Conceptualization: T.M., M.Y., Y.Z, Preparation: Y.F., R.Z., Z.W., T.Z., Experimental implementation: T.M., R.J., T.N., Data analysis: T.M., M.Y., Y.Z., Y.F., Manuscript writing: T.M., M.Y., Y.Z., R.J., T.N.

\section*{Competing Interests Statement}
We declare no competing interest.

\section*{Data and code availability}
The python script used for data analysis in this manuscript is available from the authors upon request.

\bibliographystyle{unsrtnat}
\bibliography{references.bib}
\end{document}